\documentclass[12pt]{article}
\usepackage{latexsym,amssymb,amsmath, bm}
\usepackage[margin=0.6in]{geometry}
\usepackage{graphicx} 
\usepackage{natbib}
\usepackage{hyperref}
\usepackage{amsthm}
\usepackage{xcolor}
\usepackage{comment}
\usepackage{float}
\usepackage[utf8]{inputenc}
\usepackage[english]{babel}
\usepackage{amsfonts}
\usepackage{authblk}
\usepackage{mymath}
\usepackage{subfigure,algorithm,algorithmic}
\usepackage{lineno}
\usepackage{filecontents}
\usepackage[inline, shortlabels]{enumitem}
\usepackage[title,titletoc]{appendix}

\def\blu#1{\textcolor{blue}{#1}}

\def\Fbar{\overline{F}}
\def\Gbar{\overline{G}}
\begin{document}

\title{\bfseries Pricing cyber insurance for a large-scale network\footnote{Acknowledgments: The research is supported by the Casualty Actuarial Society and the Society of Actuaries through an individual grant competition, and the Gaea HPC facility from Northern Illinois University provided computing resources for many of the simulations.}}
  \author{Lei Hua\footnote{Department of Statistics and Actuarial Science, Northern Illinois University} \;\;\; Maochao Xu\footnote{Department of Mathematics, Illinois State University}}
\date{ }

\maketitle

\subparagraph{Abstract.} Facing the lack of cyber insurance loss data, we propose an innovative approach for pricing cyber insurance for a large-scale network based on synthetic data. The synthetic data is generated by the proposed risk spreading and recovering algorithm that allows infection and recovery events to occur sequentially, and allows dependence of random waiting time to infection for different nodes. The scale-free network framework is adopted to account for the topology uncertainty  of the random large-scale network. Extensive simulation studies are conducted to understand the risk spreading and recovering mechanism, and to uncover the most important underwriting risk factors. A case study is also presented to demonstrate that the proposed approach and algorithm can be adapted accordingly to provide reference for cyber insurance pricing.


\section{Introduction}
Over the recent years, cyber insurance has become increasingly important, as our society has relied more and more on the cyber domain for almost all aspects of our daily life and work. However, there have been only a few research works dealing with modeling and assessing cyber risks. Among obstacles that have prevented the cyber insurance market from achieving maturity, the absence of reliable actuarial data and the reluctance of revealing IT infrastructures make actuarial modeling even more challenging \citep{Betterley2017,kosub2015,Xu2019}.

Traditionally, rate making relies on actuarial tables constructed from experience studies. Unlike traditional insurance policies, cyber insurance has no standard scoring systems or actuarial tables. Cyber risks are relatively new, and there are only a very small amount of data about security breaches and losses. This difficulty may be further exacerbated by the reluctance of organizations to reveal the details of security breaches for the sake of avoiding loss of market shares, loss of reputation, etc. Pricing cyber insurance is still a challenging question although the demand has been increasing and there are insurance companies providing cyber insurance products. According to Betterley's report \citep{Betterley2017}, the insurers tend to increase premiums for larger companies, and the coverage can be limited and very expensive for companies without good cyber security protection. The main feature that distinguishes cyber risks from traditional insurance risks is that the information and communication technology resources are interconnected in a network,
 {and we are interested in understanding how interconnected risks might affect each other}. Most recently, \cite{Xu2019} proposed a general framework for modeling the infection and recovery processes of a network, while accounting for various costs and risks that might arise from those processes. This method models the cyber risk from a micro-level perspective, which works very well if the exact network structure is known. 
However, in practice it may be unfeasible to extract the exact structure of the network due to the confidentiality of IT infrastructures, scales, and complexity of real-world networks. For example, the network for financial transactions of a bank can be extremely complicated, and pricing cyber insurance based on such a network becomes impractical if one has to know the exact structure of the network. Therefore, to tackle this challenging issue, we propose a novel approach that models the cyber risk from a macro level perspective  {based on some basic specifications of the network. But one should notice that there are some other factors, such as network design principles, that may affect a network's susceptibility to cyber risks}.

The most closely related work to cyber risk from the macro level is by   \cite{Eling2018}, where they studied the cost of cyber risks by using data from the SAS OpRisk Global data. They adopted a loss distribution approach which considers the loss frequency and severity, and deployed the extreme value theory to study the loss distribution. Although this work provides a further understanding of cyber risks, the network characteristics are not considered, which we believe have significant effects on the cyber risk of a specific company. Our proposed work aims to provide further understanding on how network characteristics affect the cyber losses on the macro level.  Specifically, for a large-scale network, instead of knowing the exact structure, which is often unfeasible, one could have some basic quantities such as the numbers of nodes and edges to describe the basic features of the complex network. For underwriting purposes, it is more practical if one can consider the premiums without knowing the exact structure of the network. Also, the physical network structure is likely different than the network logical topology which describes how the data flow within the network. To this end, we employ the scale-free network structures for the mechanism of generating the underling network topology. The so-called scale-free structures widely exist in the real world, and see \cite{Barabasi2016} for many real world examples of scale-free networks. Instead of knowing the exact network structure, we will only need some network statistics including the number of nodes, the number of edges, and the degree distributions for abstracting the network. One of our main tasks is to understand how those network features would affect cyber risks.

We develop a simulation-based approach for assessing potential cyber risks on a large-scale network given a set of underwriting information. The study in this paper is innovative in the following aspects: (1) the randomness of networks is accounted for, and only summary information about the network is required to conduct the simulations; (2) a more reasonable mechanism for the processes of risk infection and recovery is adapted and implemented, which allows both infection and recovery to continuously evolve over time, which is a new algorithm different than that in \cite{Xu2019}; (3) an extensive simulation study based on the implemented simulation algorithms has been conducted, uncovering insights into how factors such as network features contribute to the frequency and magnitude of cyber risks.

The paper is then organized as follows: Section \ref{sec-theapproach} introduces the proposed approach and algorithm, as well as the basic concepts of scale-free networks in Section \ref{sec-scalefree}. Then a simulation study is conducted in Section \ref{sec-sim} to understand the mechanism of risk spreading and recovering and the statistical effects on cyber risks. Particularly, in Section \ref{sec-alleffects} we report the findings about the statistical effects from various predictive variables on the cyber risks to uncover the most important underwriting risk factors. Section \ref{sec-case} demonstrates how to use the proposed approach to price a specific large-scale network. Section \ref{sec-remark} concludes the paper with further discussions.

\section{The proposed approach}
\label{sec-theapproach}
Ratemaking for cyber risks on a large-scale network is very challenging, especially considering that the network topology and risk spreading and recovering mechanism themselves can be very complicated, there are very few relevant datasets available, etc. In order to address these issues, our approach is based on synthetic data that can account for a variety of scenarios with respect to network topology, interactions between infection and recovery activities, and so on.

A loss frequency-severity model based on a large-scale network can be the following. Let $G(V,E)$ denote an undirected graph with $n$ nodes represented by $V$ and $m$ edges represented by $E$. {
A node abstracts a computer/server/working station at an appropriate resolution depending on the insuring interest, and the graph abstracts the communication network of a company. }

Let random variables $N$ and $Z$ be the number of loss events during a unit time period and the loss severity, respectively.  For a given graph $G$ and some other covariates $X_N$, the conditional number of loss events $N|(G,X_N) \sim F_{N|(G,X_N)}(\mathcal{G}, \theta)$, where $\mathcal{G}$ is the collection of attributes that are associated with the graph itself such as number of nodes and edges, and $\theta$ is the collection of the parameters for the other covariates such as average waiting time to infection and to recovery, respectively. Further, the loss severity $Z$, such as the cost of replacing a node, does not depend on the network $G$, and it may depend on some other predictive variables $X_Z$. \blu{For example, the cost of replacing a node is generally an inconsequential fraction of the cost of a breach event.}
We can write $Z|X_Z \sim F_{Z|X_Z}(\alpha)$, where $\alpha$ is the collection of parameters for the predictive variables.
Let $Z_1, \dots, Z_n$ be independent copies of $Z$, and $Z$ and $N$ are independent. Therefore, the aggregated loss for the graph $G$ can be written as
\begin{align}
S|(G,X_N,X_Z) = \sum_{i=1}^{N|(G,X_N)} Z_i|X_Z \notag.
\end{align}
and then
\begin{align}
\E[S|(X_N, X_Z)] \approx \sum_{j=1}^{|J|} \left[ \E[N|(G_j, X_N)] \cdot \E[Z|X_Z] \right]/|J|, \label{eq-ES}
\end{align}
where the $(G_1, \dots, G_{|J|})$ is a random sample of $G$, and a sampling scheme is provided by \cite{Goh2001} when $G$ is of a scale-free network, $J$ is the collection of indexes, and $|J|$ is its cardinality, i.e., the number of simulated networks. The variability of $S$ can be assessed similarly based on the simulations. The approximation in (\ref{eq-ES}) should be sufficiently accurate when $J$ is large enough. We refer to \cite{Goh2001, Chung2002, Cho2009} for sampling from a scale-free network.

Here we notice that the network topology and the risk spreading scheme are assumed only to affect the loss frequency, while the loss severity $Z$ does not depend on either the network structure or how the cyber risks spread. This assumption is reasonable as loss severity (such as the cost of labor and replacements for recovering a node) has usually been standardized and does not depend on the network itself. That being said, one of the main tasks of this paper is how to tackle the modeling task for the loss frequency $N|(G,X_N)$, for which both network structures and the cyber risk spreading scheme may play an important role.

%

\subsection{Scale-free network}
\label{sec-scalefree}
In order to account for the randomness of large-scale network structures, we consider using scale-free networks as the underling network structures. Please refer to  \cite{Barabasi2016} for a large amount of examples of scale-free networks in the real word.

When a network is too large and its exact structure is impossible to describe, such as a financial transaction network and the Internet traffic network, one often uses summary statistics to describe the structure of the network. Among those summary statistics associated with a network, the number of nodes, the number of edges, and the degree distribution of the network are the most basic ones. Roughly speaking, the degree of a node is the number of edges connected to the node, and then the degree distribution describes how the node degrees are distributed over the network. Let a random variable $K$ represent the node degrees, and then $p_k=\P[K=k]$ is the probability mass function of $K$. When $K$ follows a power-law distribution in the sense that $p_k = a k^{-\gamma}$ with constants $a >0$ and $\gamma>0$, the network is referred to as a scale-free network; for most real-world examples of scale-free networks, the range of the index $\gamma$ is between $2$ and $3$ (see, \cite{Barabasi1999, Barabasi2016}). We will focus on \blu{\emph{undirected}} scale-free networks in this paper, as such a property has been found in many different commonly-used networks that are relevant to cyber risks, such as the Internet at the router level and email networks \blu{for which two directly connected nodes can communicate with each other in both directions}. We refer to \cite{Barabasi1999,Barabasi2016} for details about scale-free networks, and
\cite{Brandes2005, Kolaczyk2009} for terminology and statistical inference about network analysis.


\subsection{A novel model and its related simulation algorithm}
Given the number of nodes, the number of edges, and the degree distribution, one can generate networks from a static scale-free random graph that preserves these given properties (see, \cite{Goh2001, Chung2002, Cho2009}). Synthetic data, and thus estimated risks associated with the randomly sampled network, can then be obtained based on a risk spreading and recovering algorithm that is to be discussed in detail in Appendix. By altering the sets of values of the basic network statistics $(n,m,\gamma)$, as well as the cost functions and the assumptions for the infection and recovery processes, one can assess various cyber risks that are associated with the random network, of which only the basic network statistics are known. The synthetic data can then be used to calibrate the model developed so that a flexible enough but simplified pricing model can be developed.

Due to the lack of available data for many of the factors that may affect the cyber risks and corresponding losses, our approach and algorithms provide a flexible tool for conducting computer experiments based on the following factors:
\begin{itemize}
\item Number of nodes of the network;
\item Number of edges among the nodes;
\item Index $\gamma$ of the scale-free network;
\item Number of initially affected nodes;
\item Dependence structure among the nodes;
\item The distribution of waiting time to infection;
\item The distribution of waiting time to recovery.
\end{itemize}
{To keep our framework more general, the infection is defined to be very broad, which can include infecting a computer  by a malware/virus, stealing sensitive information or data breach, or losing control of computers or certain software (e.g. ransomware, DDOS attacks) etc. The infection time  can be interpreted as the time needed to have/detect an infection. For example, for the data breach as an infection, the waiting time to infection can be interpreted as the time between two consecutive breaches. Similarly, the waiting time to recovery is also defined to be very broad depending on the practical scenario. For the malware infection, the recovery can be interpreted as the time needed to clean the malware, patch  security vulnerabilities, and bring the computer back to a functioning status. For the data breach, the recovery can be interpreted as the time needed to contain the breach. For example, the average time to contain a data breach was 69 days in 2018 according to Ponemon Institute (\url{https://www.ponemon.org}). }

Although one may find that {color{red}some} of these assumptions are not in general easy to specify, the simulation models at such a level of granularity provide a useful platform for creating synthetic data while accounting for the uncertainty of those factors. A feasible approach is to consider different possibilities of those assumptions and then employ the proposed model to conduct simulations. Afterwards, assessment of potential cyber risks and losses can be done by analyzing the synthetic data. It should be noticed that we aim to develop an efficient and flexible algorithm for understanding how cyber risks might evolve on a complex large-scale network. It is out of the scope of the current work for discussing how to choose those assumptions while no relevant data are available. We believe that those assumptions can be based on experts' experiences and/or experienced data ({e.g., published reports from Ponemon Institute)}. When the corresponding data is available, the statistical inference for those assumptions should be a relatively easier task. The proposed approach is suitable for various assumptions and, in what follows, we will discuss our models and algorithms assuming that the assumptions are given. After discussing the models, we will conduct simulations and case studies to understand the risk spreading and recovering mechanism, to uncover the most important underwriting risk factors, and to demonstrate how to use the proposed approach for pricing a large-scale network, of which the uncertainty of the assumptions will have been accounted for.

In \cite{Boguna2014}, a method is proposed for simulating mutually independent non-Markov stochastic processes. In this paper, we relax the independence assumption and propose a new method of simulating dependent non-Markov stochastic processes.

 {The technical details of the proposed model, and the algorithm for simulations are included in the Appendix.  { In the following}, we briefly summarize the basic idea of the algorithm. { Given a network with nodes (e.g., computers)  and links  (e.g., cables and/or wireless communication)}, the proposed simulation algorithm  includes two types of stochastic processes: (1) { Recovery process. Each infected node can be recovered according to a stochastic recovery process.}  (2) { Infection process. Each healthy node can be infected via the vulnerable  link via a stochastic infection process.}   The status of each node, being either healthy or infected, is determined by the interaction of these two types of processes. The proposed algorithm allows all the involved processes to  { evolve during a certain time period}, and allows the node status to change from one to the other.  {  Based on the algorithm, we can simulate the evolution of infection over the network,} that is, how the infection is spreading and/or recovering over a network during a time period. Because the assumptions of the stochastic processes as well as the specifications of the network may affect how the risks are evolving, we can use the algorithm to simulate different scenarios and generate synthetic data to study risk factors.}

\section{Simulation study}
\label{sec-sim}
In this section, we conduct extensive simulation studies to understand how the network characteristics  affect the cyber risks. In Section \ref{sec-explore}, we first conduct an exploratory analysis on a network with a fixed number of nodes and edges to focus on understanding the risk spreading and recovering mechanism. Then in Section \ref{sec-alleffects}, we conduct a formal statistical analysis on the effects of various factors including the number of nodes,  the number of edges, etc..

\subsection{Exploratory analysis}
\label{sec-explore}
We first carry out simulations to study the source of variability of cyber risks based on networks with a given number of nodes and edges. Due to its flexible shape and tail behavior, Weibull distribution in the following form is chosen for both the random time to recovery and the random time to infection.
$$\Fbar(\tau) = \exp\left\{ - (\mu \tau)^\alpha \right \}, \quad \mu, \alpha > 0.$$

As in reality, most scale-free networks have the $\gamma$ parameter in the range of $(2,3)$, we choose 2.1, 2.5, and 2.9 as three values of $\gamma$. For the purpose of simulation, without knowing the exact dependence structure of a complex network dependence, we can use a multivariate Gaussian copula in order to gain some basic ideas about how dependence affects overall cyber risks. The following is the cumulative distribution function (cdf) of a $d$-dimensional Gaussian copula.
\begin{align}
C(u_1, \dots, u_d) = \Phi_d(\Phi^{-1}(u_1),\dots, \Phi^{-1}(u_d), \Sigma), \notag
\end{align}
where $\Sigma$ is the correlation matrix of the $d$-dimensional standard multivariate normal cdf $\Phi_d$, and $\Phi$ is the cdf of the standard univariate normal distribution. The $D_k$ in Equation (\ref{eq-lam-ik}) can then be written as
\begin{align}
D_k(u_1,\dots,u_d)=\Phi_{d|k}(\Phi^{-1}(u_1),\dots, \Phi^{-1}(u_d); \mu_{d|k}, \Sigma_{d|k}), \notag
\end{align}
where $\Phi_{d|k}$ is the cdf of the multivariate normal distribution with mean $\mu_{d|k} = (\rho_{ik}; i \neq k)^{\top}\cdot \Phi^{-1}(u_k)$ and correlation matrix $\Sigma_{d|k}$, and one can use the R function \texttt{partial.r\{psych\}} to get the partial correlation matrix. Because the dependence is about the random waiting time to infection on the nodes that are connected directly to the infected node, it is reasonable to assume that there is only positive dependence. Therefore, in what follows, we assume positive dependence in the Gaussian copula to make the large-scale simulation more tractable. The proposed algorithm was employed to conduct the simulation analysis. Here we assume that initially there is $1$ infected node, and the correlation coefficient is assumed to be $0.5$. These assumptions will be relaxed to be more flexible when we study the statistical effects from various factors in Section \ref{sec-alleffects}. Table \ref{tab-sim1} contains the simulation settings and the corresponding results for 5 different representative cases.

Based on Cases A, B, and C from Table \ref{tab-sim1}, the values of $\gamma$ do not have a strong effect on the cyber risks in terms of the average total service-down per node and per month, denoted as node$\times$month, and the average total number of recovered nodes. However, based on \blu{Cases C}, D, and E, both the random time to infection and the random time to recovery dramatically affect the cyber risks.
In particular, when the time to recovery is not much smaller than that for infection, the cyber risks can be greatly increased.

\begin{table}[H]
	\centering
	\begin{tabular}{|c|c|c|c|c|c|c|c|c||c|c||c|c|}
			\hline
		Case &	$n$ & $m$ & $\gamma$ & $im$ & $iv$   & $rm$ & $rv$ & $ns$ & $itm$ & $itsd$ &  $rnm$  &  $rnsd$  \\
			\hline
		A &	50  & 200 & 2.1      &  1 & 1 &  0.25 & 0.25 & 800   & 0.61& 1.76& 2.72& 5.20   \\
		B &	50  & 200 & 2.9      &  1 & 1 &  0.25 & 0.25 & 800   & 0.66& 1.73& 2.89& 5.17   \\
		C &	50  & 200 & 2.5      &  1 & 1 &  0.25 & 0.25 & 800   & 0.56 &1.51 &2.61& 4.60   \\
		\hline
		D &	50  & 200 & 2.5      &  1 & 1 &  0.5 & 0.5 & 800   &17.08  &50.15  &38.30 &105.46   \\
		E &	50  & 200 & 2.5      &  1 & 1 &  1 & 1 & 800   & 266.48 &195.54 &268.97 &197.12   \\
			\hline
 	\end{tabular}
 \caption{Effects of network topology on risk assessments.
$n$: number of nodes,
$m$: number of edges,
$\gamma$: index of scale-free networks,
$im$: mean of waiting time to infection,
$iv$: variance of waiting time to infection,
$rm$: mean of waiting time to recovery,
$rv$: variance of waiting time to recovery,
$ns$: number of simulation for each set of assumptions,
$itm$: sample mean of the total service-down time of the nodes in the unit of node$\times$month,
$itsd$: sample standard deviation of the total service-down time of the nodes in the unit of node$\times$month,
$rnm$: sample mean of the total number of nodes that have been recovered,
$rnsd$: sample standard deviation of the total number of nodes that have been recovered.
}
\label{tab-sim1}	
\end{table}

\begin{figure}[htp]
\begin{center}
\includegraphics[width=1\textwidth]{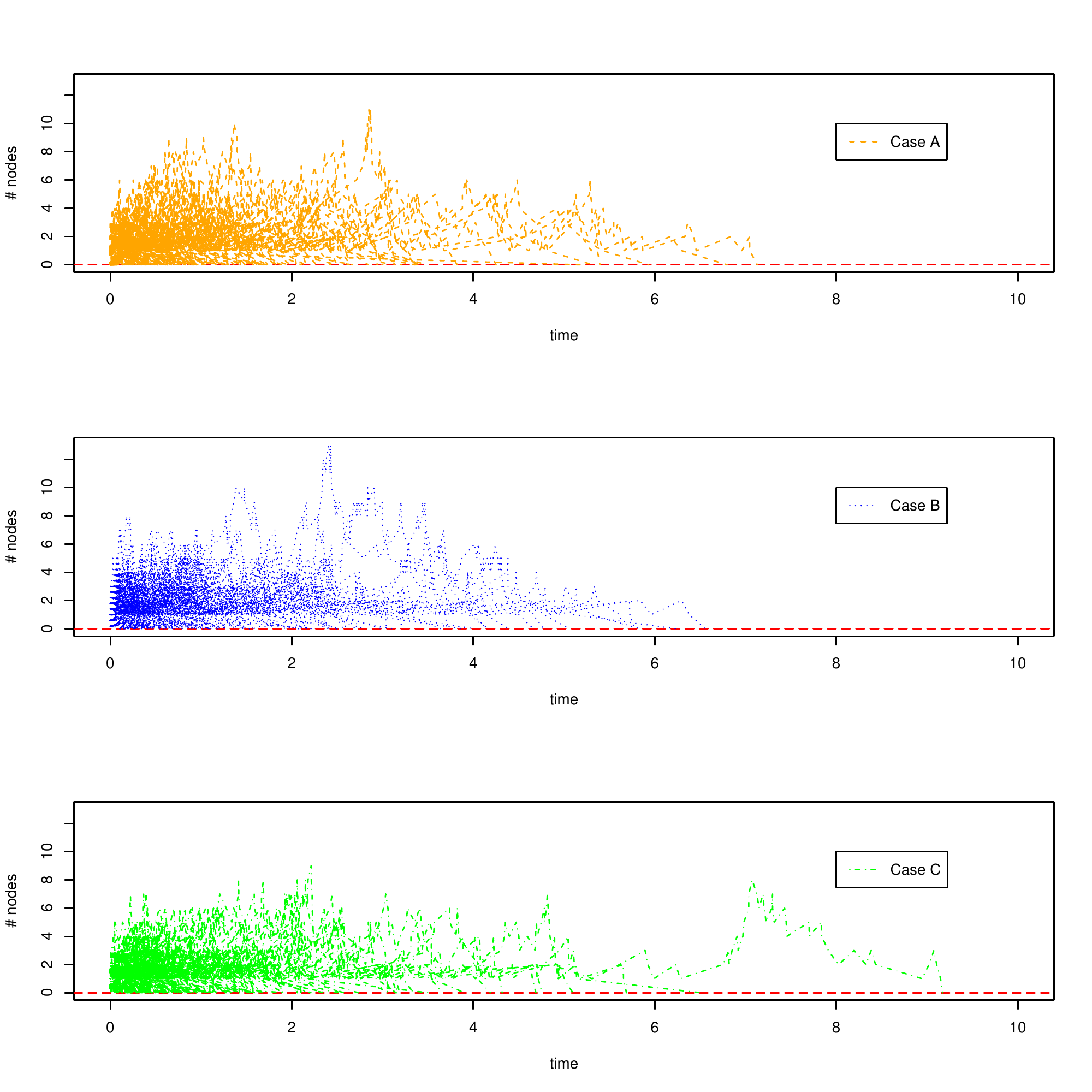}
\end{center}
\caption{Numbers of infected nodes along time} \label{fig-ABC}
\end{figure}

Figures \ref{fig-ABC} illustrates the trajectories of the numbers of infected nodes for Cases A, B and C, respectively with 800 randomly developed trajectories for each case. Note that with the same number of nodes and the same number of edges, etc., different $\gamma \in (2,3)$ for the scale-free network does not lead to significantly different development patterns of the infected nodes. This observation is consistent to the numbers reported in Table \ref{tab-sim1} as well.

\begin{figure}[htp]
\begin{center}
\includegraphics[width=1\textwidth]{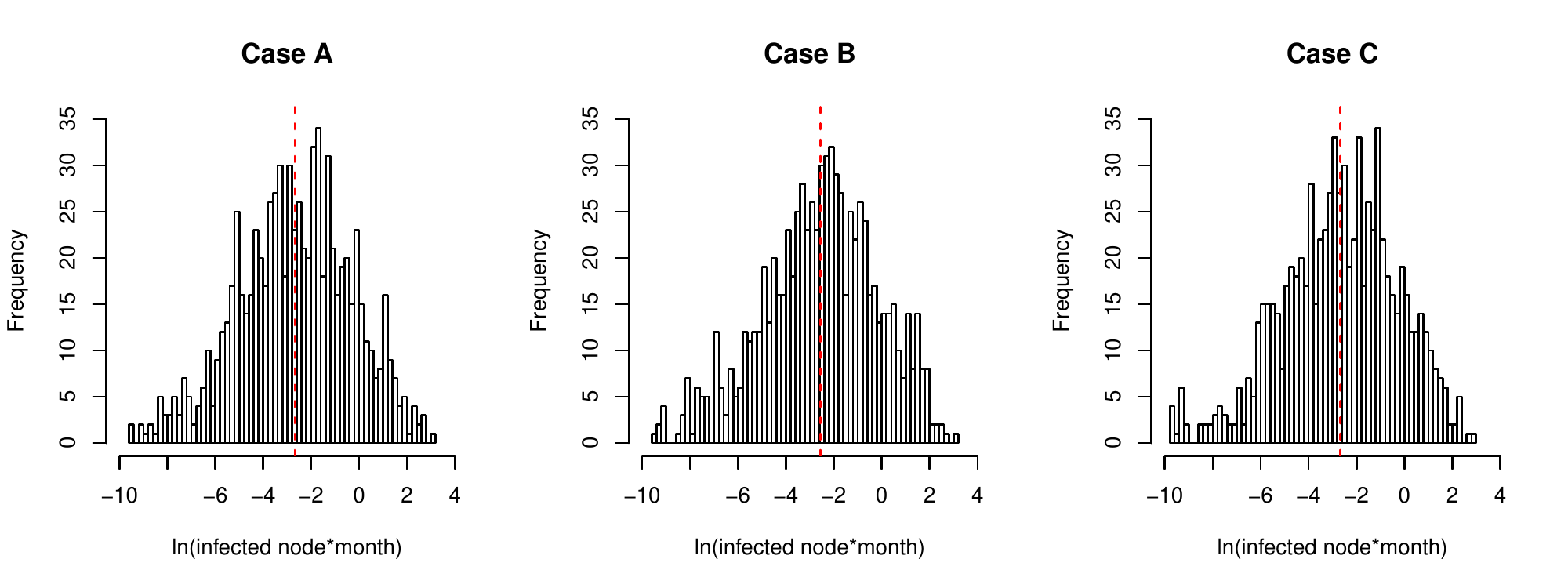}
\end{center}
\caption{Histogram for the natural logarithm of accumulated infected node$\times$month} \label{fig-ABC-hist}
\end{figure}

Figure \ref{fig-ABC-hist} contains the histograms of the natural logarithm of the accumulated infected node$\times$month for Cases A, B, and C, respectively. The vertical red dashed lines are the corresponding sample means. Again, the patterns are quite similar although the $\gamma$'s are different.

What we can conclude from the above comparisons is that given that the other factors are the same, the $\gamma$ parameter for the scale-free network may not affect the overall cyber risks significantly. In Section \ref{sec-alleffects}, a further study with various factors such as different number of nodes and different number of edges involved will suggest that $\gamma$ is not statistically significant in affecting the cyber risks.

\begin{figure}[htp]
\begin{center}
\includegraphics[width=.8\textwidth]{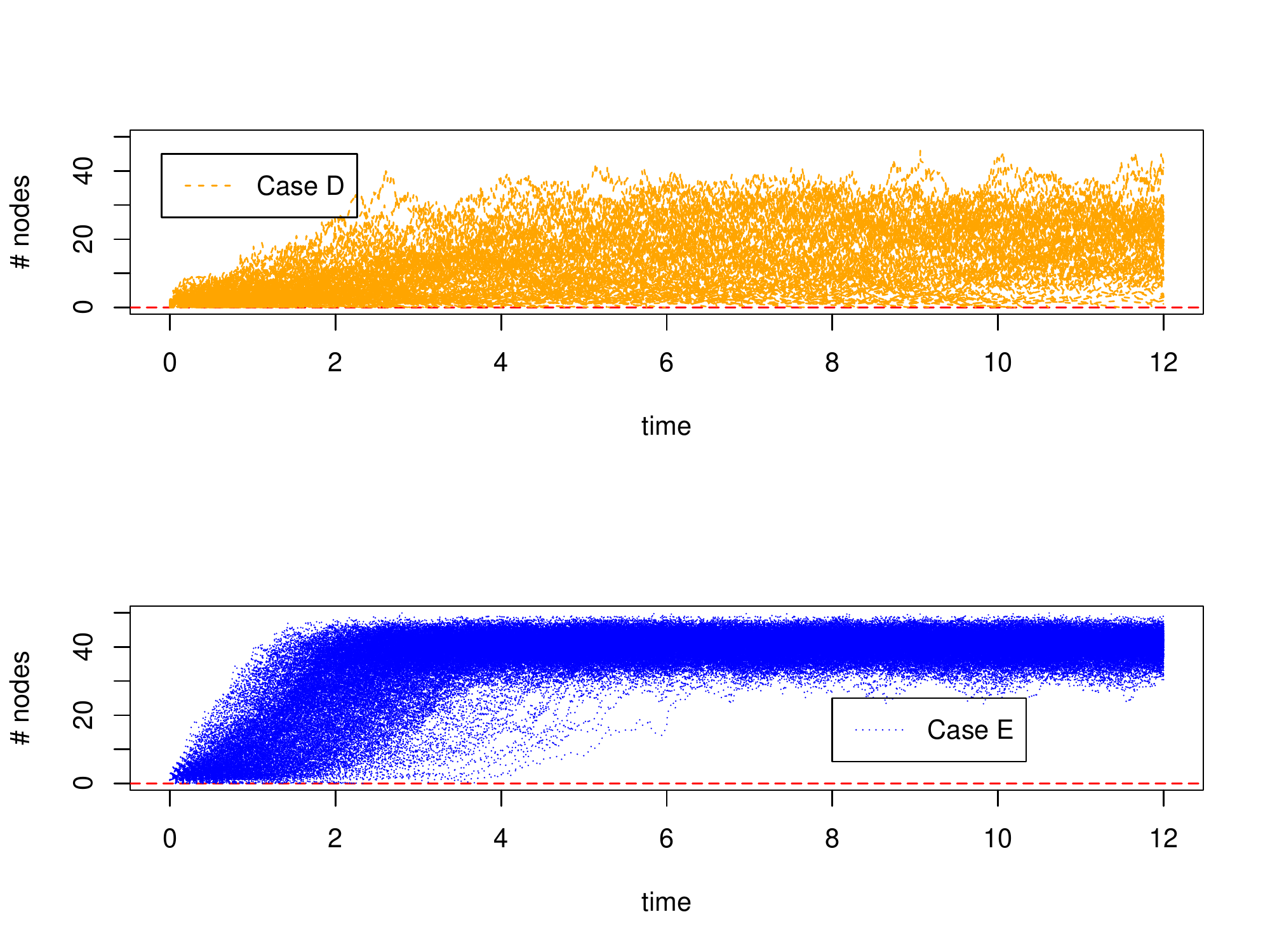}
\end{center}
\caption{Numbers of infected nodes along time} \label{fig-DE}
\end{figure}

Figure \ref{fig-DE} compares the trajectories of the numbers of infected nodes for Cases D and E, respectively with 800 randomly developed trajectories for each case. Note that with the same number of nodes and the same number of edges, etc., the random time to infection and the random time to recovery play a very important role in affecting the process of risk spreading and recovering. Unlike the cases in Figure \ref{fig-ABC}, Cases D and E here have a relatively slower speed for recovery compared to that for infection. Therefore, the number of infected nodes significantly increases with time. When the recovery speed is slow enough, such as in Case E, it is most likely that the number of infected nodes is increasing dramatically until almost all are infected. In this case, the whole network will have a very slim chance of a complete recovery.

\begin{figure}[htp]
\begin{center}
\includegraphics[width=0.8\textwidth]{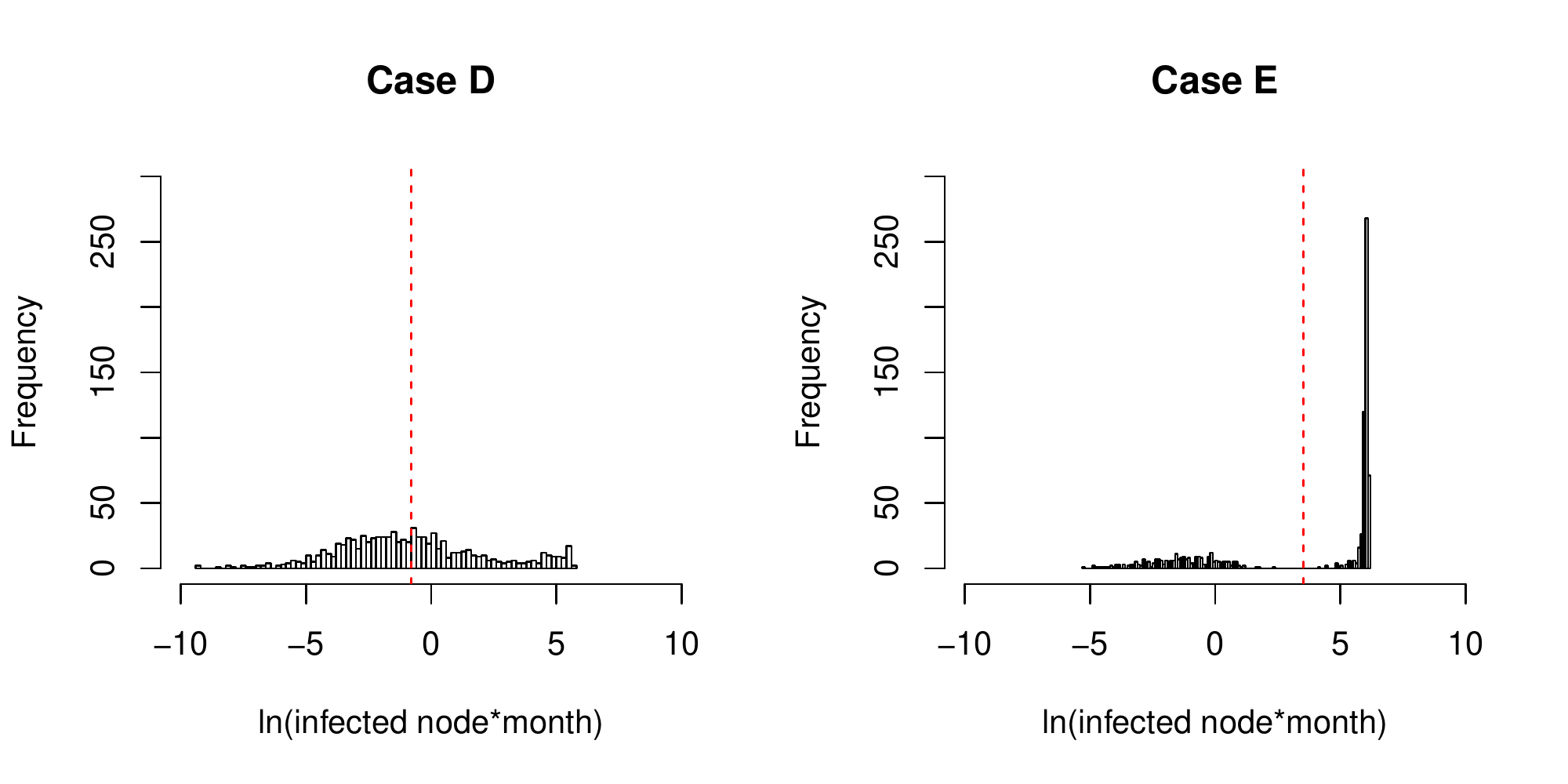}
\end{center}
\caption{Histogram for the natural logarithm of accumulated infected node$\times$month} \label{fig-DE-hist}
\end{figure}

Figure \ref{fig-DE-hist} shows the histograms of the natural logarithm of the accumulated infected node$\times$month for Cases D and E, respectively. The vertical red dashed lines are the corresponding sample means. It is interesting to note that there tend to be two modes for the distributions of the accumulated infected node$\times$month when the time to recovery is getting closer to the time to infection, which is clearer in Case E. The possible reason is that in the network system, the recovery and infection are competing with each other, and when one of these two competing factors dominates, the whole system tends to be dominated by such a factor.

\clearpage

\subsection{Statistical effects of network and risk spreading features}
\label{sec-alleffects}
In this section, we formally study the statistical effects from various factors on cyber risks based on synthetic data. The purpose of assessing the statistical effects from various factors is to identify those important variables that have to be included in the candidate models for pricing cyber risk insurance. Our approach is to randomly choose the values of the parameters associated with the model, and then simulations are conducted to generate synthetic data for further statistical analysis. We have generated a random sample of size $800$ within the time frame $T=12$. The following predictive variables are considered:
\begin{itemize}
\setlength\itemsep{-0.4em}
\item \texttt{par\_cop:} the common correlation coefficient $\rho$ of the Gaussian copula (from $0.1$ to $0.9$);
\item \texttt{mean\_rec:} the population mean of the Weibull distribution for the random waiting time to recovery (from 0.1 to 1);
\item \texttt{var\_rec:} the population variance of the Weibull distribution for the random waiting time to recovery (from 0.1 to 1);
\item \texttt{mean\_inf:} the population mean of the Weibull distribution for the random waiting time to infection (from 0.1 to 6);
\item \texttt{var\_inf:} the population variance of the Weibull distribution for the random waiting time to infection (from 0.1 to 6);
\item \texttt{Nnode:} number of nodes (from $20$ to $100$);
\item \texttt{Nedge:} number of edges (from $80$ to $400$);
\item \texttt{Gam:} the $\gamma$ parameter for the scale-free network (from 2 to 3);
\item \texttt{Ninf0:} initial number of infected nodes (from 1 to 5).
\end{itemize}

\begin{figure}[htp]
\begin{center}
\includegraphics[width=0.8\textwidth]{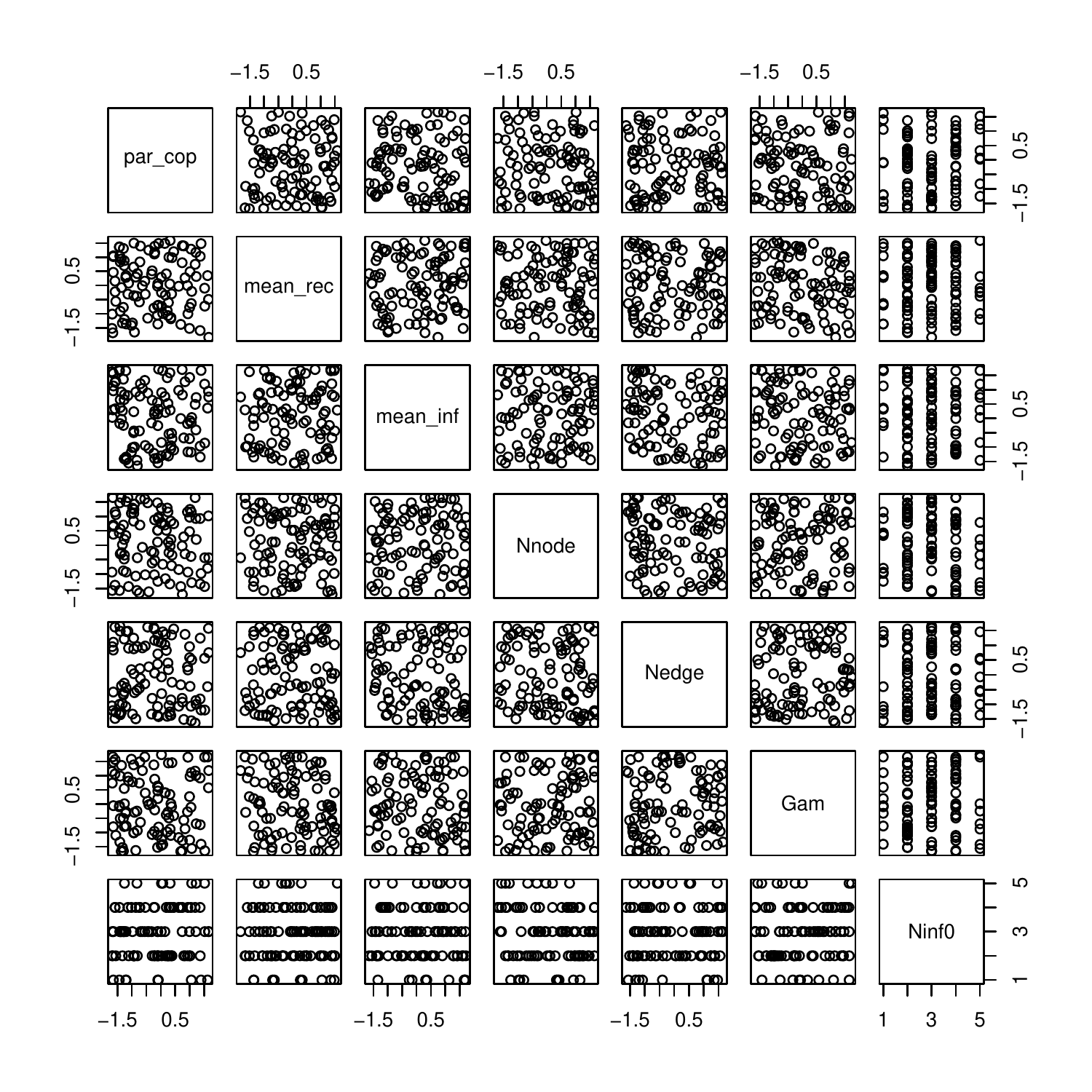}
\end{center}
\caption{Scatter plots of standarlized covariates} \label{fig-covariates}
\end{figure}

In order to check that the predictive variables are roughly balanced and mutually independent, we first standardize them except for the variable \texttt{Ninf0}, and then draw the scatter plots. Figure \ref{fig-covariates} contains the scatter plots of randomly chosen $100$ points for these covariates, and there seem to be no multicolinearity and the data are roughly balanced.

There are two main response variables that are of interest: \texttt{Tinf}, the accumulated infected node$\times$month, and \texttt{Nrec}, the accumulated number of recovered nodes, and these two variables directly affect the amount of losses. For example, \texttt{Tinf} may lead to losses due to the shut down of services, and \texttt{Nrec} is directly associated with repair and replacement costs \blu{which however are not generally significant costs of a cyber insurance claim.}

Based on the ranges of the predictive variables, we use some strictly increasing transformations such as the logit and natural logarithm functions to make the ranges of the variables be $(-\oo, \oo)$ to have relatively stable numerical performance when regression analyses are conducted with computer software. 
Please refer to Tables \ref{tab-est} and \ref{tab-est-Nrec} for the details on how such transformations were employed, and their corresponding estimates.

\subsection{Effects on the accumulated infected nodes and time}
For the response variable \texttt{Tinf}, we have tried several different models such as linear models and generalized linear models based on the gamma distribution and the inverse Gaussian distribution. After trying the Box-Cox transformation with the help of the R function \texttt{boxcox()}, we find that the optimal value of $\lambda$ of the Box-Cox transformation is $-0.02$ for the data. Since this value is very close to $0$, we can simply transform the response variable by the natural logarithm. Afterwards, a multiple linear regression can be used to assess the effects of the predictive variables. The following linear regression model was chosen to assess the effects of the predictive variables.

\begin{verbatim}
      ln(Tinf) ~ logit(par_cop)+ln(mean_rec)+ln(mean_inf)+ln(var_rec)+ln(var_inf)+
                 ln(Nnode)+ln(Nedge)+logit(Gam-2)+Ninf0
\end{verbatim}

Tables \ref{tab-est} and \ref{tab-ANOVA} are the corresponding estimates and the ANOVA table, respectively. There are several interesting findings suggested by the analysis. First, from Table \ref{tab-ANOVA} we find that the means of the recovery time and the infection time contribute most to the variability of the accumulated infected node$\times$month, and the initial number of infected nodes plays a relatively important role.
However, after controlling for the other variables, the number of nodes and the number of edges in the network do not affect the response variable as much as the aforementioned leading factors do. This suggests that the assumptions about the average recovery and infection time are the most critical ones, while the network features such as the number of nodes, the number of edges, and the $\gamma$ parameter are relatively less important. Second, the dependence parameter does play a role here although it is not very important. Under such a simulation setting, a more mutually dependent network tends to lead to a relatively smaller \texttt{Tinf}; see Table \ref{tab-est}.
A reasonable interpretation is related to Equation (\ref{eq-Phi_tau}), where relatively stronger upper tail dependence might lead to a larger value of $\Phi(\tau|\{t_{j_s}, r_j\})$, thus a larger probability that no infection occurs in the next time period of $\tau$. However, in general there are no deterministic results about the direction of the effect from dependence.
Third, the directions of the effects from the significant predictive variables are consistent to our intuition based on Table \ref{tab-est}. For example, a larger time to recovery and a smaller time to infection tend to increase \texttt{Tinf}, and a larger number of initially infected nodes tends to increase \texttt{Tinf} as well. Fourth, as we have already observed from Section \ref{sec-sim}, the $\gamma$ parameter for the scale-free network is not significant at all after controlling all the other variables, which removes our concerns about the influence from the index of scale-free networks.

\begin{table}[ht]
\centering
\begin{tabular}{|r|r|r|r|r|l|}
  \hline
 & Estimate & SE & t-value & p-value &   \\
  \hline
(Intercept) & 0.2972 & 1.0700 & 0.2777 & 0.781293 &   \\
  logit(par\_cop) & -0.2512 & 0.0607 & -4.1402 & 0.000038 & *** \\
  log(mean\_rec) & 1.9968 & 0.1208 & 16.5258 & $<$ 2.2e-16 & *** \\
  log(mean\_inf) & -1.3011 & 0.0857 & -15.1792 & $<$ 2.2e-16 & *** \\
  log(var\_rec) & -0.4671 & 0.1121 & -4.1650 & 0.000035 & *** \\
  log(var\_inf) & 0.1017 & 0.1090 & 0.9325 & 0.351360 &   \\
  log(Nnode) & 0.4421 & 0.1689 & 2.6169 & 0.009044 & ** \\
  log(Nedge) & -0.1778 & 0.1508 & -1.1786 & 0.238931 &   \\
  logit(Gam - 2) & -0.0607 & 0.0388 & -1.5646 & 0.118086 &   \\
  Ninf02 & 1.2327 & 0.2345 & 5.2576 & 1.9e-7 & *** \\
  Ninf03 & 2.0192 & 0.2376 & 8.4981 & $<$ 2.2e-16 & *** \\
  Ninf04 & 2.2612 & 0.2349 & 9.6271 & $<$ 2.2e-16 & *** \\
  Ninf05 & 2.4142 & 0.2760 & 8.7481 & $<$ 2.2e-16 & *** \\
   \hline
\end{tabular}
\caption{Estimates of the regression coefficients for \texttt{Tinf}, and the reference level of the categorical variable is \texttt{Ninf0} $=1$}
\label{tab-est}
\end{table}

\begin{table}[htp]
\centering
\begin{tabular}{|r|r|r|r|r|r|r|l|}
  \hline
 & DF & SumSq & RSS & AIC & F-value & p-value &   \\
  \hline
$<$none$>$ &  &  & 2797.1 & 1027.4 &  &  &   \\
  logit(par\_cop) & 1 & 60.92 & 2858.1 & 1042.6 & 17.1417 & 0.000038 & *** \\
  log(mean\_rec) & 1 & 970.65 & 3767.8 & 1263.7 & 273.1025 & $<$ 2.2e-16 & *** \\
  log(mean\_inf) & 1 & 818.92 & 3616.1 & 1230.8 & 230.4096 & $<$ 2.2e-16 & *** \\
  log(var\_rec) & 1 & 61.66 & 2858.8 & 1042.8 & 17.3473 & 0.000035 & *** \\
  log(var\_inf) & 1 & 3.09 & 2800.2 & 1026.3 & 0.8696 & 0.351360 &   \\
  log(Nnode) & 1 & 24.34 & 2821.5 & 1032.3 & 6.8480 & 0.009044 & ** \\
  log(Nedge) & 1 & 4.94 & 2802.1 & 1026.8 & 1.3890 & 0.238931 &   \\
  logit(Gam - 2) & 1 & 8.70 & 2805.8 & 1027.9 & 2.4479 & 0.118086 &   \\
  Ninf0 & 4 & 448.45 & 3245.6 & 1138.4 & 31.5439 & $<$ 2.2e-16 & *** \\
   \hline
\end{tabular}
\caption{ANOVA analysis for \texttt{Tinf}}
\label{tab-ANOVA}
\end{table}

\subsection{Effects on the accumulated recovered nodes}
For the response variable \texttt{Nrec}, we have tried generalized linear models based on the Poisson and the negative binomial distributions, respectively. A test of overdispersion proposed in \cite{Cameron1990} indicates that the negative binomial is more suitable, for which we employed the R function \texttt{dispersiontest()} in the \texttt{AER} package. The following negative binomial model was chosen to assess the effects of the predictive variables.

\begin{verbatim}
      Nrec ~ logit(par_cop)+ln(mean_rec)+ln(mean_inf)+ln(var_rec)+ln(var_inf)+
             ln(Nnode)+ln(Nedge)+logit(Gam-2)+Ninf0
\end{verbatim}

Tables \ref{tab-est-Nrec} and \ref{tab-type3} are the corresponding estimates and the Type III analysis, respectively. The patterns of the effects from the predictive variables on \texttt{Tinf} are quite similar to those on \texttt{Nrec}. For instance, based on Table \ref{tab-type3}, average infection time and recovery time contribute most to the variability of the accumulated number of recovered nodes, and after controlling the other variables, the number of nodes and the number of edges in the network do not affect \texttt{Nrec} as much as the leading factors. This again suggests that the assumptions about the average recovery and infection time are the most critical ones. Moreover, the $\gamma$ parameter again does not play a significant role. Like the case for \texttt{Tinf}, the dependence parameter also plays a significant role with stronger dependence leading to fewer recovered nodes.

\begin{table}[htp]
\centering
\begin{tabular}{|r|r|r|r|r|l|}
  \hline
 & Estimate & SE & t-value & p-value &   \\
  \hline
(Intercept) & 2.0887 & 0.7393 & 2.8251 & 0.004726 & ** \\
  logit(par\_cop) & -0.3592 & 0.0419 & -8.5817 & $<$ 2.2e-16 & *** \\
  log(mean\_rec) & 0.8462 & 0.0850 & 9.9593 & $<$ 2.2e-16 & *** \\
  log(mean\_inf) & -2.3463 & 0.0583 & -40.2360 & $<$ 2.2e-16 & *** \\
  log(var\_rec) & -0.2237 & 0.0765 & -2.9253 & 0.003441 & ** \\
  log(var\_inf) & 0.4216 & 0.0747 & 5.6400 & 1.7e-8 & *** \\
  log(Nnode) & 0.0997 & 0.1161 & 0.8590 & 0.390343 &   \\
  log(Nedge) & 0.3485 & 0.1041 & 3.3474 & 0.000816 & *** \\
  logit(Gam - 2) & -0.0063 & 0.0268 & -0.2367 & 0.812924 &   \\
  Ninf02 & 0.0889 & 0.1640 & 0.5421 & 0.587720 &   \\
  Ninf03 & 0.5082 & 0.1654 & 3.0731 & 0.002118 & ** \\
  Ninf04 & 0.7928 & 0.1631 & 4.8621 & 0.000001 & *** \\
  Ninf05 & 0.6543 & 0.1898 & 3.4478 & 0.000565 & *** \\
   \hline
\end{tabular}
\caption{Estimates of the regression coefficients for \texttt{Nrec}, and the reference level of the categorical variable is \texttt{Ninf0} $=1$}
\label{tab-est-Nrec}
\end{table}

\begin{table}[htp]
\centering
\begin{tabular}{|r|r|r|r|r|r|l|}
  \hline
 & DF & Deviance & AIC & LRT & p-value &   \\
  \hline
$<$none$>$ &  & 894.0 & 6182.0 &  &  &   \\
  logit(par\_cop) & 1 & 965.0 & 6251.0 & 70.970 & $<$ 2.2e-16 & *** \\
  log(mean\_rec) & 1 & 988.0 & 6273.0 & 93.530 & $<$ 2.2e-16 & *** \\
  log(mean\_inf) & 1 & 2259.0 & 7545.0 & 1364.690 & $<$ 2.2e-16 & *** \\
  log(var\_rec) & 1 & 903.0 & 6188.0 & 8.480 & 0.0036 & ** \\
  log(var\_inf) & 1 & 921.0 & 6206.0 & 26.520 & 2.6e-7 & *** \\
  log(Nnode) & 1 & 895.0 & 6181.0 & 0.640 & 0.4226 &   \\
  log(Nedge) & 1 & 905.0 & 6190.0 & 10.090 & 0.0015 & ** \\
  logit(Gam - 2) & 1 & 895.0 & 6180.0 & 0.060 & 0.8123 &   \\
  Ninf0 & 4 & 935.0 & 6215.0 & 40.960 & 2.7e-8 & *** \\
   \hline
\end{tabular}
\caption{Type III analysis for \texttt{Nrec}}
\label{tab-type3}
\end{table}

\section{Case study}
\label{sec-case}
In Section \ref{sec-sim}, we used a relatively smaller network to study the statistical effects of various factors. However, when the network of interest is much larger, one needs to conduct simulations accordingly based on the actual specification of interest. In this section, we consider a particular case of much larger networks, and demonstrate how the potential cyber risks can be assessed under the proposed framework, and how to use the proposed model for pricing cyber insurance for a large-scale network.

The following are the assumptions considered for this particular example.
\begin{enumerate}
\item The target market is large networks with 100 to 5000 nodes and the number of edges from 400 to 20000 edges.
\item The network is a scale-free network, and the $\lambda$ parameter is from 2 to 3.
\item The policy term is 12 months $(T=12)$.
\item The waiting time to both infection and recovery follows a Weibull distribution.
Moreover, the average and the standard deviation of the random time to recovery are from $0.1$ to $1$ month, respectively, and those of the random time to infection are from $0.1$ to $\sqrt{6}$ months, respectively.
\item The dependence structure among the random waiting time to infection of the nodes that are linked to the same affected node is a Gaussian copula with the pairwise correlation coefficient $\rho$, which can take a value from $0.1$ to $0.9$.
\item The cost of recovering a node is $\eta$ and the loss of service interruption per node$\times$month is $\omega$.
\end{enumerate}

Note that the above assumptions cover a very wide range of different cases considering the following aspects: (a) The numbers of nodes and edges are quite flexible; (b) Only the distribution family for the random waiting time is assumed, and the mean and variance can be chosen based on the real applications; (c) The dependence structure covers a wide range of positive dependence.

%

Similarly to the study in Section \ref{sec-alleffects}, we first generate synthetic data based on the above assumptions, and the sample size is 1200. We have checked that there are no multicolinearity of the covariates and the data are balanced across different variables.

For \texttt{Tinf}, Box-Cox transformation is applied to transform the response variable \texttt{Tinf} to roughly follow a normal distribution. A multiple linear regression model is then fitted with all the covariates being included, and the ANOVA table is Table \ref{tab-type3-case-0}. It suggests that the following variables do not contribute significantly to the model: \texttt{par\_cop}, \texttt{var\_inf}, \texttt{Nnode}, \texttt{Nedge}, and \texttt{Gam}.

Then we exclude those insignificant variables except \texttt{var\_inf} as by all means we need to include it as otherwise the exact distribution for the random waiting time to infection cannot be specified. Then the estimated parameter $\hat{\lambda} = 0.1818182$ for the Box-Cox transformation for the candidate model, and the estimates and the ANOVA table are Table \ref{tab-case-est} and Table \ref{tab-case-type3}, respectively.

\begin{table}[ht]
\centering
\begin{tabular}{|r|r|r|r|r|r|r|l|}
  \hline
 & DF & SumSq & RSS & AIC & F-value & p-value &   \\
  \hline
$<$none$>$ &  &  & 1638.8 & 400.0 &  &  &   \\
  logit(par\_cop) & 1 & 4.58 & 1643.4 & 401.3 & 3.3153 & 0.0689 & . \\
  log(mean\_rec) & 1 & 802.33 & 2441.1 & 876.2 & 581.1431 & $<$ 2.2e-16 & *** \\
  log(mean\_inf) & 1 & 37.41 & 1676.2 & 425.0 & 27.1004 & 2.3e-7 & *** \\
  log(var\_rec) & 1 & 41.73 & 1680.5 & 428.1 & 30.2284 & 4.7e-8 & *** \\
  log(var\_inf) & 1 & 0.00 & 1638.8 & 398.0 & 0.0034 & 0.9538 &   \\
  log(Nnode) & 1 & 0.06 & 1638.8 & 398.0 & 0.0412 & 0.8392 &   \\
  log(Nedge) & 1 & 0.83 & 1639.6 & 398.6 & 0.6037 & 0.4373 &   \\
  logit(Gam - 2) & 1 & 0.00 & 1638.8 & 398.0 & 0.0001 & 0.9943 &   \\
  Ninf0 & 4 & 754.20 & 2393.0 & 846.3 & 136.5708 & $<$ 2.2e-16 & *** \\
   \hline
\end{tabular}
\caption{ANOVA table for \texttt{Tinf}: case study with all variables included}
\label{tab-type3-case-0}
\end{table}

\begin{table}[ht]
\centering
\begin{tabular}{|r|r|r|r|r|r|r|l|}
  \hline
 & DF & SumSq & RSS & AIC & F-value & p-value &   \\
  \hline
$<$none$>$ &  &  & 1644.0 & 395.8 &  &  &   \\
  log(mean\_rec) & 1 & 815.84 & 2459.8 & 877.3 & 591.0427 & $<$ 2.2e-16 & *** \\
  log(mean\_inf) & 1 & 42.22 & 1686.2 & 424.2 & 30.5871 & 3.9e-8 & *** \\
  log(var\_rec) & 1 & 40.52 & 1684.5 & 423.0 & 29.3540 & 7.3e-8 & *** \\
  log(var\_inf) & 1 & 0.00 & 1644.0 & 393.8 & 0.0028 & 0.9577 &   \\
  Ninf0 & 4 & 760.96 & 2405.0 & 844.3 & 137.8215 & $<$ 2.2e-16 & *** \\
   \hline
\end{tabular}
\caption{ANOVA table for \texttt{Tinf}: case study with the candidate model}
\label{tab-case-type3}
\end{table}

\begin{table}[ht]
\centering
\begin{tabular}{|r|r|r|r|r|l|}
  \hline
 & Estimate & SE & t-value & p-value &   \\
  \hline
(Intercept) & 0.0171 & 0.1192 & 0.1435 & 0.885955 &   \\
  log(mean\_rec) & 1.4782 & 0.0608 & 24.3114 & $<$ 2.2e-16 & *** \\
  log(mean\_inf) & -0.3380 & 0.0611 & -5.5306 & 3.9e-8 & *** \\
  log(var\_rec) & -0.3167 & 0.0585 & -5.4179 & 7.3e-8 & *** \\
  log(var\_inf) & 0.0027 & 0.0505 & 0.0530 & 0.957732 &   \\
  Ninf02 & 0.7874 & 0.1030 & 7.6421 & 4.4e-14 & *** \\
  Ninf03 & 1.4568 & 0.1070 & 13.6147 & $<$ 2.2e-16 & *** \\
  Ninf04 & 1.8960 & 0.1054 & 17.9819 & $<$ 2.2e-16 & *** \\
  Ninf05 & 2.1917 & 0.1081 & 20.2765 & $<$ 2.2e-16 & *** \\
   \hline
\end{tabular}
\caption{Estimates for \texttt{Tinf}: case study with the candidate model}
\label{tab-case-est}
\end{table}

For \texttt{Nrec}, Poisson regression and the negative binomial regression have been compared, and the overdispersion test suggests that the negative binomial regression model is more suitable. An initial model with all the variables included leads to the Type III analysis shown in Table \ref{tab-case-type3-Nrec}. It can be seen that \texttt{par\_cop}, \texttt{var\_rec}, and \texttt{Nnode} do not contribute to the deviance significantly; although the variables \texttt{Nedge} and \texttt{Gam} appear to be statistically significant, they contribute only marginally to the deviance. Based on the above consideration, we choose the same set of predictive variables for the candidate model for the response variable \texttt{Nrec} as that for \texttt{Tinf}, and the model is as follows:

\begin{verbatim}
           Nrec ~ ln(mean_rec)+ln(mean_inf)+ln(var_rec)+ln(var_inf)+Ninf0.
\end{verbatim}

Then the estimated regression parameters and Type III analysis are reported in Table \ref{tab-case-est-Nrec-candidate} and Table \ref{tab-case-type3-Nrec-candidate}, respectively.

\begin{table}[ht]
\centering
\begin{tabular}{|r|r|r|r|r|r|l|}
  \hline
 & DF & Deviance & AIC & LRT & p-value &   \\
  \hline
$<$none$>$ &  & 946.0 & 5581.0 &  &  &   \\
  logit(par\_cop) & 1 & 946.0 & 5579.0 & 0.240 & 0.6276 &   \\
  log(mean\_rec) & 1 & 991.0 & 5624.0 & 44.930 & 2.0e-11 & *** \\
  log(mean\_inf) & 1 & 1474.0 & 6107.0 & 528.510 & $<$ 2.2e-16 & *** \\
  log(var\_rec) & 1 & 946.0 & 5579.0 & 0.390 & 0.5322 &   \\
  log(var\_inf) & 1 & 979.0 & 5612.0 & 32.900 & 9.7e-9 & *** \\
  log(Nnode) & 1 & 948.0 & 5581.0 & 2.480 & 0.1156 &   \\
  log(Nedge) & 1 & 951.0 & 5584.0 & 5.030 & 0.0249 & * \\
  logit(Gam - 2) & 1 & 950.0 & 5583.0 & 4.320 & 0.0376 & * \\
  Ninf0 & 4 & 1403.0 & 6030.0 & 457.090 & $<$ 2.2e-16 & *** \\
   \hline
\end{tabular}
\caption{Type III analysis for \texttt{Nrec}: case study with all variables included}
\label{tab-case-type3-Nrec}
\end{table}

\begin{table}[ht]
\centering
\begin{tabular}{|r|r|r|r|r|r|l|}
  \hline
 & DF & Deviance & AIC & LRT & p-value &   \\
  \hline
$<$none$>$ &  & 946.0 & 5584.0 &  &  &   \\
  log(mean\_rec) & 1 & 988.0 & 5624.0 & 42.300 & 7.8e-11 & *** \\
  log(mean\_inf) & 1 & 1488.0 & 6123.0 & 541.450 & $<$ 2.2e-16 & *** \\
  log(var\_rec) & 1 & 947.0 & 5582.0 & 0.590 & 0.4439 &   \\
  log(var\_inf) & 1 & 981.0 & 5616.0 & 34.530 & 4.2e-9 & *** \\
  Ninf0 & 4 & 1400.0 & 6029.0 & 453.460 & $<$ 2.2e-16 & *** \\
   \hline
\end{tabular}
\caption{Type III analysis for \texttt{Nrec}: case study with candidate model}
\label{tab-case-type3-Nrec-candidate}
\end{table}

\begin{table}[ht]
\centering
\begin{tabular}{|r|r|r|r|r|l|}
  \hline
 & Estimate & SE & t-value & p-value &   \\
  \hline
(Intercept) & 1.6069 & 0.0782 & 20.5405 & $<$ 2.2e-16 & *** \\
  log(mean\_rec) & 0.2595 & 0.0395 & 6.5670 & 5.1e-11 & *** \\
  log(mean\_inf) & -0.8514 & 0.0370 & -23.0366 & $<$ 2.2e-16 & *** \\
  log(var\_rec) & -0.0291 & 0.0374 & -0.7783 & 0.436419 &   \\
  log(var\_inf) & 0.1917 & 0.0325 & 5.8995 & 3.6e-9 & *** \\
  Ninf02 & 0.3607 & 0.0744 & 4.8499 & 0.000001 & *** \\
  Ninf03 & 0.7532 & 0.0739 & 10.1854 & $<$ 2.2e-16 & *** \\
  Ninf04 & 1.0551 & 0.0708 & 14.9104 & $<$ 2.2e-16 & *** \\
  Ninf05 & 1.3013 & 0.0715 & 18.1869 & $<$ 2.2e-16 & *** \\
   \hline
\end{tabular}
\caption{Estimates for \texttt{Nrec}: case study with candidate model}
\label{tab-case-est-Nrec-candidate}
\end{table}

 Now that we have Table \ref{tab-case-est} and Table \ref{tab-case-est-Nrec-candidate}, the estimates for the respective candidate model for \texttt{Tinf} and \texttt{Nrec} respectively, the expected total loss can then be estimated as
\begin{align}
\E[S] = \omega \cdot \E[\texttt{Tinf}]  + \eta \cdot \E[\texttt{Nrec}]. \label{eq-predictedS}
\end{align}

Knowing the values of $\omega$ and $\eta$, the predicted total loss $\hat{S}$ can be trivially calculated based on Equation (\ref{eq-predictedS}).
Table \ref{tab-pred-TinfNrec} contains the predicted \texttt{Tinf} and \texttt{Nrec}, and their standard errors. The last two columns of the table are based on the case where the values of $\omega$ and $\eta$ are specified, for which a detailed discussion follows.

\begin{table}[ht]
\centering
\begin{tabular}{|r|r|r|r|r|cc|cc||cc|}
  \hline
mean\_rec & mean\_inf & var\_rec & var\_inf & Ninf0 & $\widehat{\texttt{Tinf}}$ & s.e. & $\widehat{\texttt{Nrec}}$ & s.e. & $\hat{S}$ (k usd) & s.e. \\
  \hline
0.2 & 1 & 0.2 & 1 & 1 & 0.105 & 0.020 & 3.442 & 0.284 & 74.075 & 6.665 \\
  0.4 & 1 & 0.2 & 1 & 1 & 0.408 & 0.057 & 4.121 & 0.319 & 102.807 & 9.205 \\
  0.2 & 2 & 0.2 & 1 & 1 & 0.073 & 0.013 & 1.908 & 0.151 & 41.788 & 3.699 \\
  0.4 & 2 & 0.2 & 1 & 1 & 0.307 & 0.040 & 2.284 & 0.164 & 61.047 & 5.257 \\
  0.2 & 1 & 0.2 & 1 & 2 & 0.306 & 0.047 & 4.937 & 0.393 & 114.061 & 10.215 \\
  0.4 & 1 & 0.2 & 1 & 2 & 0.960 & 0.114 & 5.910 & 0.440 & 166.231 & 14.485 \\
  0.2 & 2 & 0.2 & 1 & 2 & 0.227 & 0.034 & 2.737 & 0.207 & 66.094 & 5.811 \\
  0.4 & 2 & 0.2 & 1 & 2 & 0.755 & 0.081 & 3.276 & 0.222 & 103.249 & 8.500 \\
  0.2 & 1 & 0.2 & 1 & 3 & 0.663 & 0.090 & 7.311 & 0.571 & 179.394 & 15.934 \\
  0.4 & 1 & 0.2 & 1 & 3 & 1.814 & 0.196 & 8.752 & 0.645 & 265.754 & 22.678 \\
  0.2 & 2 & 0.2 & 1 & 3 & 0.512 & 0.066 & 4.052 & 0.297 & 106.664 & 9.263 \\
  0.4 & 2 & 0.2 & 1 & 3 & 1.464 & 0.143 & 4.851 & 0.322 & 170.233 & 13.595 \\
  0.2 & 1 & 0.2 & 1 & 4 & 1.045 & 0.131 & 9.887 & 0.761 & 249.976 & 21.755 \\
  0.4 & 1 & 0.2 & 1 & 4 & 2.655 & 0.265 & 11.835 & 0.851 & 369.451 & 30.274 \\
  0.2 & 2 & 0.2 & 1 & 4 & 0.824 & 0.099 & 5.480 & 0.398 & 150.793 & 12.926 \\
  0.4 & 2 & 0.2 & 1 & 4 & 2.174 & 0.199 & 6.560 & 0.426 & 239.906 & 18.489 \\
  0.2 & 1 & 0.2 & 1 & 5 & 1.390 & 0.171 & 12.647 & 0.990 & 322.462 & 28.335 \\
  0.4 & 1 & 0.2 & 1 & 5 & 3.382 & 0.337 & 15.140 & 1.117 & 471.913 & 39.197 \\
  0.2 & 2 & 0.2 & 1 & 5 & 1.110 & 0.130 & 7.010 & 0.509 & 195.697 & 16.647 \\
  0.4 & 2 & 0.2 & 1 & 5 & 2.794 & 0.253 & 8.391 & 0.549 & 307.545 & 23.619 \\
   \hline
\end{tabular}
\caption{Predicted Tinf and Nrec, and the total loss for a special case}
\label{tab-pred-TinfNrec}
\end{table}

\begin{figure}[htp]
\begin{center}
\includegraphics[width=0.6\textwidth]{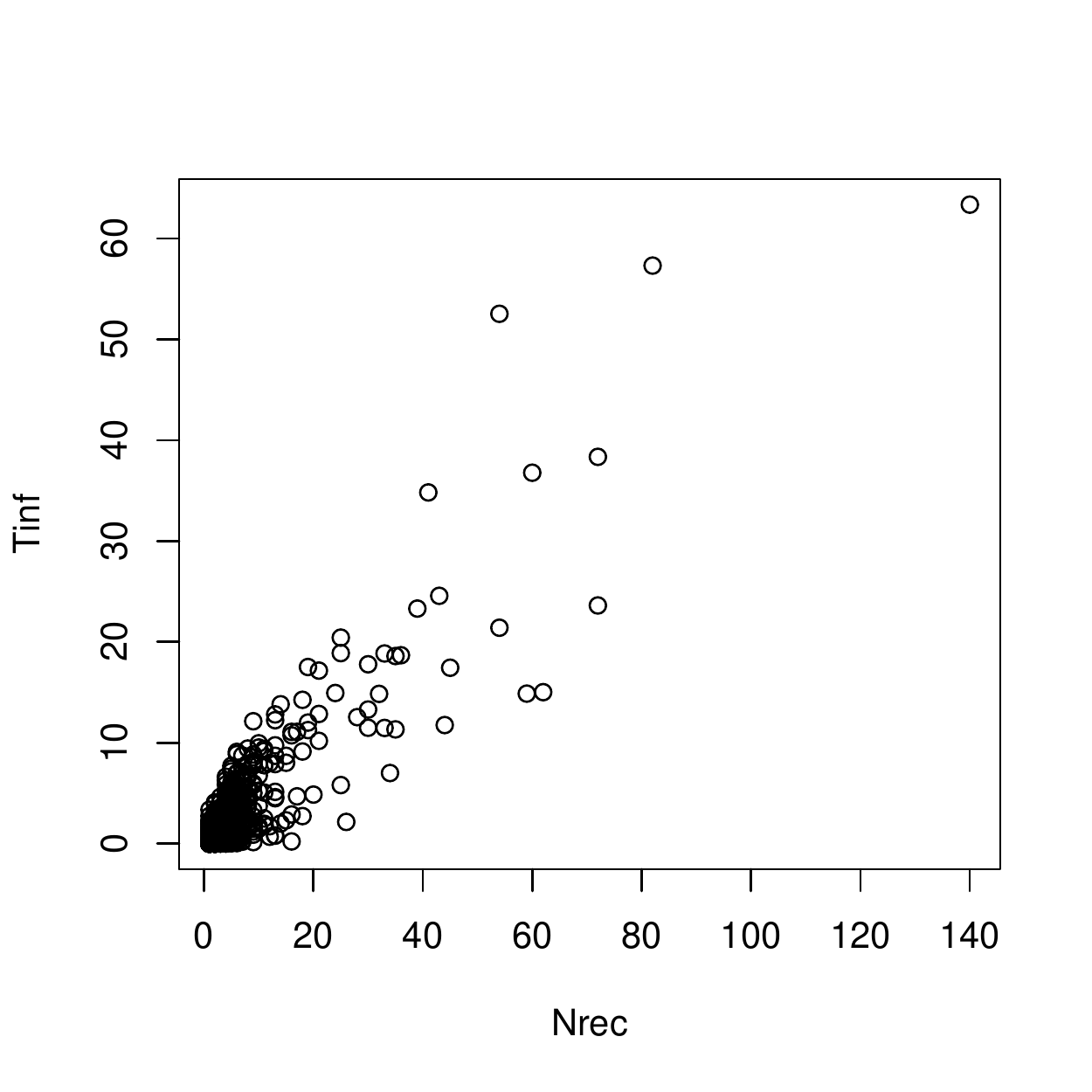}
\end{center}
\caption{Scatter plots for \texttt{Tinf} and \texttt{Nrec} in the case study} \label{fig-TinfNrec}
\end{figure}

In order to assess the standard error of $\hat{S}$, one needs to consider the covariance between $\widehat{\texttt{Tinf}}$ and $\widehat{\texttt{Nrec}}$. A theoretically ideal method is to model the covariance between \texttt{Tinf} and \texttt{Nrec} conditioning on different values of the predictive variables. However, the scatter plot (see Figure \ref{fig-TinfNrec}) between them indicates that the dependence is very strong, and the calculated correlation coefficient is about $0.88$. Under such a situation, a conservative method can be used to assess the standard error of $\hat{S}$. That is, consider an upper bound of the covariance between $\widehat{\texttt{Tinf}}$ and $\widehat{\texttt{Nrec}}$. Therefore, by Cauchy-Schwarz inequality,
\begin{align}
\var(S) \leq \omega^2 \var(\texttt{Tinf}) + \eta^2 \var(\texttt{Nrec})+2\omega \eta \sqrt{\var(\texttt{Tinf}) \var(\texttt{Nrec}) }. \notag
\end{align}
For example, assume that $\omega = 50$k usd and $\eta = 20$k usd, then the predicted $\hat{S}$ and its standard error can be calculated as the last two columns of Table \ref{tab-pred-TinfNrec}.

Through such a case study we have some interesting findings. First, the number of nodes and the number of edges may not be that important on determining the overall cyber risks under the proposed framework and the assumptions. This may be because the numbers of nodes and edges are already large enough to allow cyber risks to spread and recover freely. The most important factors are how fast the risks are spreading, how fast the infected nodes are recovered, and how many nodes have initially been infected. The estimates of the overall cyber risks can be very sensitive to the changes of those factors. Second, although the dependence structure may affect the overall risks, it plays a marginal role. There are two aspects we should notice about the role of dependence structures: On the one hand, the dependence structure is assumed only on the random waiting time to infection, and in reality dependence among other factors may exist; on the other hand, there are no deterministic results about the direction of the effects from the dependence structure; it might increase or decrease the overall risks under the current framework.


\section{Concluding remarks}
\label{sec-remark}
Facing the shortage of credible data for pricing cyber insurance for a large-scale network, we  propose a novel approach based on synthetic data that are generated from a flexible risk spreading and recovering algorithm. The proposed algorithm is able to handle the following scenarios: it allows infection and recovery to occur sequentially over the term, and it allows dependence among the random waiting time to infection. The proposed approach provides a more practical way for assessing the cyber risks of a large-scale network, while only requiring a reasonably small set of underwriting information.

After a large amount of simulation studies and a case study, we find that pricing cyber insurance for a large-scale network the following factors are the most important ones that deserve higher attention and priority: random waiting time to infection and to recovery, and the number of infected nodes at the beginning of the term. {Of these three factors, time to recovery could be estimated based on the underwriting inquiry or the empirical data, and the number of infected nodes at the beginning of the term could be related to some score system that evaluates overall vulnerability of the network as a whole to the outside networks.} The other factors, such as the number of nodes, the number of edges, and the degree of the scale-free network, are much less important compared to those leading factors. This conclusion at least holds within the conducted studies and assumptions, and for that to hold under other scenarios, we think that the number of initially affected nodes should be much smaller than the number of the nodes and edges of the whole network, to allow risk spreading and recovery without constraint coming from the network topology itself.

A potential limitation of the proposed risk spreading and recovery algorithm is that it does not account for self-infection,  {such as infections from USB flash drive attacks}, an infection without linking to any other nodes.
 {Under an intranet environment or some other internal networks with restricted access to USB flash drives and any other outside networks,} the issue becomes minimal. Otherwise, {practically one can
either artificially add one node connected by one edge to each existing node for the simulations, or use the aforementioned overall vulnerability score system to factor the self-infection risks in. }


\clearpage

 {\section*{Appendix}}
\label{appendix}

 {\subsection*{Algorithms for simulating cyber risks spreading and recovering}}

First of all, we define an \emph{active link} as the link that satisfies the following two conditions: (1) exactly two nodes are connected by the link; (2) exactly one of the two nodes is infected. For a given infected node $j$, its $k$-th active link is indexed by $j_k$. Assume that there are $M$ infected nodes at time $t$, and the set of the indexes of the infected nodes is denoted as $\mathcal{I}$. Each infected node generates a recovery process with the random waiting time to recovery represented by $R_{j}$ with the survival function $\overline{G}_j(r), j \in \mathcal{I}$.
Each active link generates an infection process with the random waiting time to infection represented by $T_{j_k}$ with the survival function $\overline{F}_{j_k}(t)$, $j=1,\ldots, M$, and for each given $j$, $k = 1, \dots, k_{j}$, where $k_j$ is the number of active links associated with the infected node $j$; define $\mathcal{A}_j$ the set of active links associated with the $j$th infected node. Let $N$ represent the total number of active links, and then $N = \sum_{j=1}^M k_j$.

A reasonable assumption is that the infection processes from the same infected node are assumed to be dependent; that is, $T_{j_1}, \dots, T_{j_{k_j}}$ are dependent and their dependence structure can be modeled by a copula $C$. Specifically, assume that the infected node $j$ launches attacks (through active links) to its susceptible neighbors $\{j_{1},\ldots,j_{k_j}\}$. Then $(T_{j_1}, T_{j_2},\ldots,T_{j_{k_j}})$ is assumed to have the following joint survival function
$$\Fbar_j(t_{j_1},\ldots,t_{j_{k_j}})=C_j\left(\Fbar_{j_1}\left(t_{j_1}\right),\ldots,\Fbar_{j_{k_j}}(t_{j_{k_j}})\right).$$
Then, the joint survival function for all the $T_{j_k}$s is
\begin{align}
\Fbar(t_{j_s}, j \in \mathcal{I}, s \in \mathcal{A}_j) = \prod_{j=1}^M \Fbar_j(t_{j_{1}},\ldots,t_{j_{k_j}})= \prod_{j=1}^M C_j\left(\Fbar_{j_1}\left(t_{j_1}\right),\ldots,\Fbar_{j_j}(t_{j_{k_j}})\right).
\end{align}

\begin{figure}[htp]
\begin{center}
\includegraphics[width=.45\textwidth]{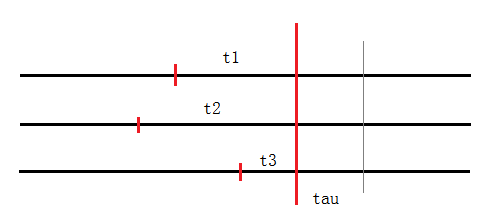}
\end{center}
\caption{Illustration about elapsed time $\{t_{j_s}\}$ and $\tau$} \label{fig-time}
\end{figure}

Given the time $\{t_{j_k}\}, j \in \mathcal{I}, k \in \mathcal{A}_j$ have elapsed since these active links have been activated, and the time $\{r_j\}, j \in \mathcal{I}$ have elapsed since these nodes have been infected, the probability that the next infection corresponds to process $i_k$ and will occur at time $t_{i_k}+\tau$ can be represented as follows, where $D_{i_k}(u_1, \dots, u_{k_i}) := \partial C_i(u_1, \dots, u_{k_i})/ \partial u_k$.

\begin{align}
& \phi(\tau,i_k| t_{j_s}, r_j, j \in \mathcal{I}, s \in \mathcal{A}_j) \notag \\
& = \lim_{\Delta \ra 0}\P\left(T_{i_k}\in(t_{i_k}+\tau,t_{i_k}+\tau + \Delta),T_{j_s}>t_{j_s}+\tau, j \neq i, s \neq k, R_w > r_w + \tau,  w \in \mathcal{I} \right. \notag \\
& \quad \quad \quad \quad \left. | T_{j_s}>t_{j_s}, R_j > r_j, j \in \mathcal{I}, s \in \mathcal{A}_j \right)/\Delta \notag \\
& = \frac{\lim_{\Delta \ra 0}\P\left(T_{i_k}\in(t_{i_k}+\tau,t_{i_k}+\tau + \Delta),T_{j_s}>t_{j_s}+\tau, j \neq i, s \neq k, R_w > r_w + \tau,  w \in \mathcal{I} \right)/\Delta}{\P\left(T_{j_s}>t_{j_s}, R_j > r_j, j \in \mathcal{I}, s \in \mathcal{A}_j\right)} \notag \\
& = D_{i_k}\left(\Fbar_{i_1}(t_{i_1}+\tau),\ldots,\Fbar_{i_{k_i}}(t_{i_{k_i}}+\tau)\right)f_{i_k}(t_{i_k}+\tau) \notag \\
& \quad\times \frac{\prod_{j \in \mathcal{I}} \Gbar_j(r_{j} + \tau)}{ \prod_{j \in \mathcal{I}} \Gbar_j(r_{j})} \times \frac{\prod_{j\in \mathcal{I} \setminus \{i \} } C_j\left(\Fbar_{j_1}(t_{j_1}+\tau ),\ldots,\Fbar_{j_{k_j}}(t_{j_{k_j}}+\tau)\right)}{\prod_{j \in \mathcal{I}} C_j\left(\Fbar_{j_1}(t_{j_1}),\ldots,\Fbar_{j_{k_j}}(t_{j_{k_j}})\right)} \notag \\
& =: \frac{D_{i_k}\left(\Fbar_{i_1}(t_{i_1}+\tau),\ldots,\Fbar_{i_{k_i}}(t_{i_{k_i}}+\tau)\right)f_{i_k}(t_{i_k}+\tau)} {C_i\left(\Fbar_{i_1}\left(t_{i_1}+\tau\right),\ldots,\Fbar_{i_{k_i}}(t_{i_{k_i}}+\tau)\right)} \times \Phi(\tau| t_{j_s}, r_j, j \in \mathcal{I}, s \in \mathcal{A}_j) \notag \\
& =: \lambda_{i_k}\left(\Fbar_{i_1}(t_{i_1}+\tau),\ldots,\Fbar_{i_{k_i}}(t_{i_{k_i}}+\tau)\right) \times  \Phi(\tau| t_{j_s}, r_j, j \in \mathcal{I}, s \in \mathcal{A}_j) \label{eq-lam-ik}
\end{align}
where
\begin{align}
\Phi(\tau|\{t_{j_s}, r_j\}) & := \Phi(\tau| t_{j_s}, r_j, j \in \mathcal{I}, s \in \mathcal{A}_j) \notag \\
& = \frac{\prod_{j \in \mathcal{I}} \Gbar_j(r_{j} + \tau)}{ \prod_{j \in \mathcal{I}} \Gbar_j(r_{j})} \times \frac{\prod_{j\in \mathcal{I} } C_j\left(\Fbar_{j_1}\left(t_{j_1}+\tau\right),\ldots,\Fbar_{j_{k_j}}(t_{j_{k_j}}+\tau)\right)}{\prod_{j \in \mathcal{I}} C_j\left(\Fbar_{j_1}(t_{j_1}),\ldots,\Fbar_{j_{k_j}}(t_{j_{k_j}})\right)}, \label{eq-Phi_tau}
\end{align}
and it represents the probability that no infection occurs in the next time period of length $\tau$, given that the time $\{t_{j_s}\}$s have elapsed for those active links, and the time $\{r_j\}$s have elapsed for those infected nodes.

Similarly, the probability that the next recovery corresponds to the node $i$ and will occur at time $r_{i}+\tau$ can be represented as follows
\begin{align}
& \phi(\tau,i| t_{j_s}, r_j, j \in \mathcal{I}, s \in \mathcal{A}_j) = \frac{g_i(r_i+\tau)}{\Gbar_i(r_i+\tau)} \times \Phi(\tau|\{t_{j_s}, r_j\})=: \Lambda_i(r_i+\tau) \times \Phi(\tau|\{t_{j_s}, r_j\}), \label{eq-Lam-i}
\end{align}
where $g_i$ is the density function of $G_i$. Given the occurrence time $\tau$, the probability that the next occurring infection event belongs to process $i_k$ is
\begin{align}
		p_{i_k}=\frac{\lambda_{i_k}}{\sum_{j \in \mathcal{I}, s \in \mathcal{A}_j} \lambda_{j_s} + \sum_{j \in \mathcal{I}}\Lambda_j}, \label{eq-p}
\end{align}		
and the probability that the next recovery event belongs to node $i$ is
\begin{align}
		p_{i}=\frac{\Lambda_{i}}{\sum_{j \in \mathcal{I}, s \in \mathcal{A}_j} \lambda_{j_s} + \sum_{j \in \mathcal{I}}\Lambda_j}. \label{eq-pnode}
\end{align}	

Note that the $C_j$ here is the copula for the active links associated with the infected node $j$. When there is only one active link, the copula is not required but the marginal survival function should be kept; that is, $C_j(\Fbar_{j_1}(t_{j_1}+\tau)) \equiv \Fbar_{j_1}(t_{j_1}+\tau)$. For the situation when an infected node, say $j$, does not have any active link, that is, $\mathcal{A}_j = \emptyset$, we let $C_j \equiv 1$. The following Algorithm \ref{alg-exact} illustrates how cyber risks spread and recover continuously on a network.

\begin{algorithm}[!http]
\caption{Simulation based on exact networks}
\label{alg-exact}
INPUT: Network topology A; time span of simulation study $T$; the iteration limit; initial states of nodes; initial elapsed time $t_0$; dependence structures $C(|\boldsymbol{\theta})$; active link infection distribution;
recovery distribution.
\begin{algorithmic}[1]
\FOR{$i = 1$  \TO  iteration limit}
\WHILE{$t\leq T$ and the number of infected nodes $>0$}
\STATE{Generate a $u$ from $\mathcal{U}(0,1)$, and then solve $\tau$ according to (\ref{eq-Phi_tau}) $\Phi(\tau|\{t_{j_s}, r_j\})=u$; update the system time by adding the derived time $\tau$.}\label{alg1-3}
\STATE{For the active links and the infected nodes, calculate the $\lambda_{i_k}$ and $\Lambda_i$, respectively, based on the equations (\ref{eq-lam-ik}) and (\ref{eq-Lam-i}), and thus the probabilities of (\ref{eq-p}) and (\ref{eq-pnode}).}
\STATE{Based on the probabilities derived above, randomly sample the index corresponding to which event -- either infection or recovery -- that occurs.}
\IF{Infection occurs}
\STATE{Change the corresponding infected node status from $0$ to $1$; change the status of the corresponding active link from $1$ to $0$; assign the newly infected node with elapsed time 0.}
\ELSE
\STATE{Change the corresponding recovered node status from $1$ to $0$.}
\ENDIF
\STATE{Update the matrix representing the active links; update the elapsed time for each active link and each infected node, respectively.}
\STATE{Update the current time $t\leftarrow t+t_1$}
\RETURN{$t$, node status, matrix for the active links, elapsed time for the active links and for the infected nodes.}
\ENDWHILE
\ENDFOR
\end{algorithmic}
OUTPUT: Node status, matrix for the active links, elapsed time for the active links and for the infected nodes right after each infection or recovery event.
\end{algorithm}

The reason for Item \ref{alg1-3} of Algorithm \ref{alg-exact} is that $\Phi(\tau|\{t_{j_s}, r_j\})$ represents the probability that no event occurs during the next time length $\tau$.  Therefore, $1-\Phi(\tau|\{t_{j_s, r_j}\})$ represents the probability that at least one event occurring during the next time length $\tau$. Generate $u$ from the uniform distribution $\mathcal{U}(0,1)$, and solve $\tau$ as follows to derive the time length.
$$1-\Phi(\tau|\{t_{j_s}, r_j\})=u\sim \mathcal{U}(0,1).$$
Note that, given that an event happens, we then need to know which process it corresponds to, and this can be calculated based on Equations (\ref{eq-p}) and (\ref{eq-pnode}).

%

\medskip

\begin{figure}[htp]
\begin{center}
\includegraphics[width=1\textwidth]{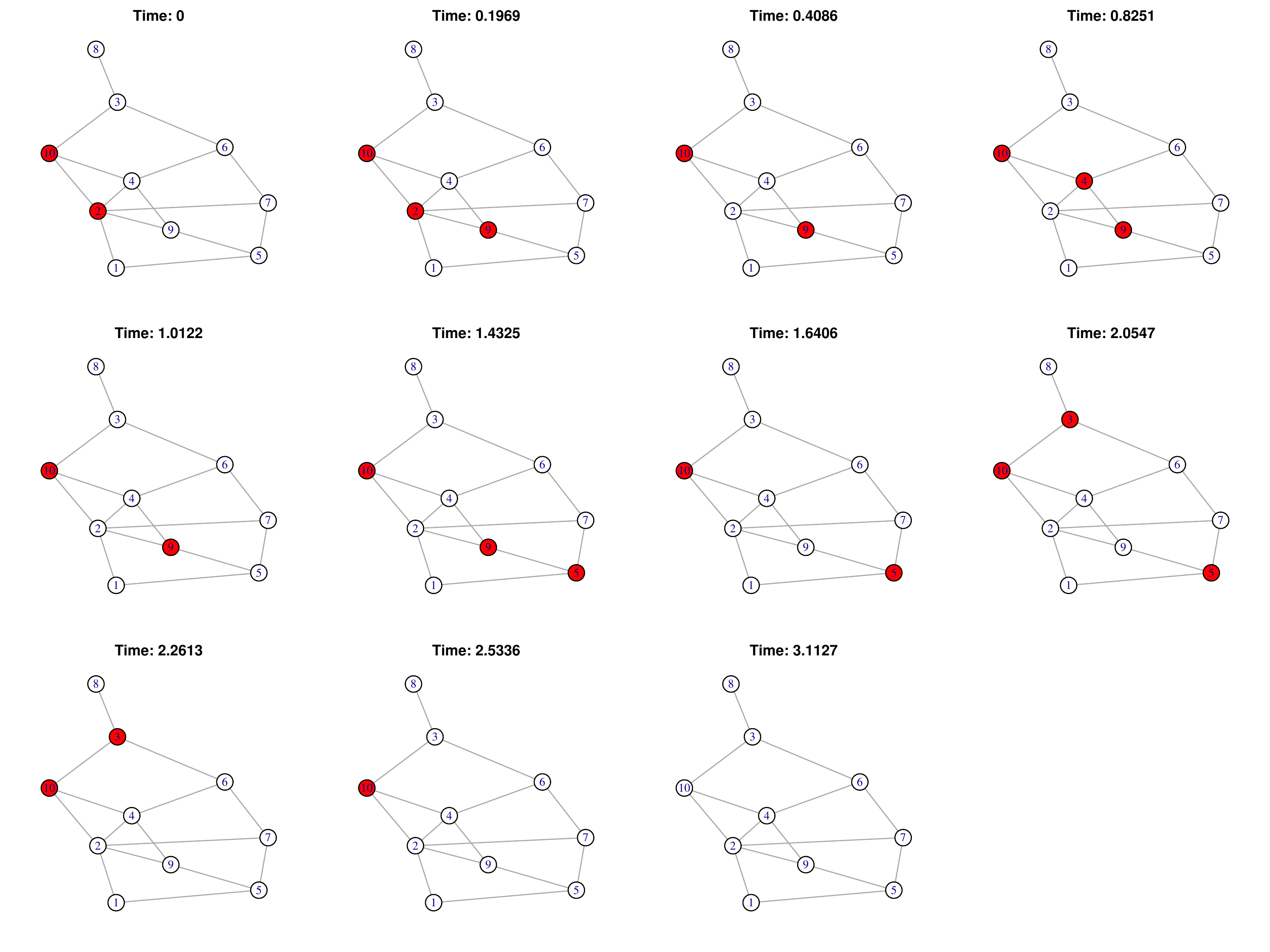}
\end{center}
\caption{Evolution of infection and recovery} \label{fig-evol}
\end{figure}

 {In Figure \ref{fig-evol}, we illustrate the interaction of the infection process and the recovery process for a network with 10 nodes and 15 edges. We let the infection propagate and the infected nodes recover by themselves until all the nodes are recovered. At each time when there is an event, either infection or recovery, the nodes in red are infected and the others are healthy. The average recovery time is assumed to be 0.8 and the average infection time is assumed to be 1. Therefore, the network tends to be completely recovered eventually, although this is not guaranteed according to the randomness of both the infection and the recovery processes.}

\end{document}